%% file: main.tex
\input{setup.tex}

\begin{document}

\input{title.tex}

\begin{strip}
    \centering
    \begin{minipage}{.8\textwidth}
            \vspace{-3em}
            \input{sections/abstract.tex}
        \hspace{2cm}
    \end{minipage}
\end{strip}

\input{sections/introduction.tex}
\input{sections/wes.tex}

\input{sections/methods.tex}

\input{sections/simulation_results.tex}

\input{sections/discussion.tex}
\input{sections/akcnowledgements.tex}
\appendix
\input{sections/app_cost.tex}

\printbibliography
\end{document}

%% file: setup.tex
\documentclass[11pt,a4paper,twocolumn]{article}
\usepackage[a4paper,margin=2cm]{geometry}

\usepackage[backend=biber,style=numeric]{biblatex}
\usepackage{algorithm}
\usepackage{algpseudocode}

\MakeRobust{\Call}

\newcommand{\Callt}[2]{\Call{#1}{\texttt{#2}}}
\newcommand{\Gets}[1]{$#1 \gets $}
\newcommand{\Getst}[1]{$\texttt{#1} \gets $}
\newcommand{\ttt}[1]{\texttt{#1}}

\algdef{SE}[SUBALG]{Indent}{EndIndent}{}{\algorithmicend\ }%
\algtext*{Indent}
\algtext*{EndIndent}

\usepackage{graphicx}
\usepackage{subcaption}
\usepackage{hyperref}
\usepackage{authblk}
\usepackage{amsmath}
\usepackage{cuted}
\usepackage{booktabs}
\usepackage{colortbl}
\usepackage[table]{xcolor}
\definecolor{lightgray}{gray}{0.9}
\usepackage{siunitx}
\usepackage{titlesec}
\titlespacing*{\section}{0pt}{4pt}{7pt}
\newcolumntype{g}{>{\columncolor{gray!15}}c}

%% file: title.tex
\title{Low Cost Bayesian Experimental Design for Quantum Frequency Estimation with Decoherence}
\vspace{-5pt}
\author[1,2,3]{Alexandra Ramôa} 
\author[1,2,3]{Luís Paulo Santos}
\author[4,5, 6]{Akihito Soeda}

\renewcommand\Affilfont{\small \itshape}

\affil[1]{International Iberian Nanotechnology Laboratory (INL), Portugal}
\affil[2]{High-Assurance Software Laboratory (HASLab), INESC TEC, Portugal}
\affil[3]{Department of Computer Science, University of Minho, Portugal}
\affil[4]{National Institute of Informatics, Chiyoda-ku, Tokyo, Japan}
\affil[5]{Department of Informatics, School of Multidisciplinary Sciences, SOKENDAI, Japan}
\affil[6]{Department of Physics, Graduate School of Science, The University of Tokyo, Japan}

\date{}

\maketitle
\vspace{-4em}

%% file: sections/abstract.tex
A two-level quantum system evolving under a time-independent Hamiltonian produces oscillatory measurement probabilities. The estimation of the associated frequency is a cornerstone problem in quantum metrology, sensing, calibration and control. In this work, we tackle this task by introducing WES: a Window Expansion Strategy for low cost adaptive Bayesian experimental design. WES employs empirical cost-reduction techniques to keep the optimization overhead low, curb scaling problems, and enable high degrees of parallelism. Unlike previous heuristics, it offers adjustable classical processing costs that determine the performance standard. As a benchmark, we analyze the performance of widely adopted heuristics, comparing them with the fundamental limits of metrology and a baseline random strategy. Numerical simulations show that WES delivers the most reliable performance and fastest learning rate, saturating the Heisenberg limit. 

%% file: sections/introduction.tex
\section{Introduction}

Frequency estimation is a crucial task in quantum metrology. It underpins the understanding of physical phenomena such as Larmor \cite{malcomlevitt2008}, Rabi \cite{merlin2021} and Ramsey \cite{ramsey1950} oscillations. Applications include calibration of superconducting qubits and magnetic field sensing \cite{santagati2018, craigie2021, msc}. Other closely related quantum problems are iterative phase estimation \cite{kitaev1995, Kitaev_2002, Wiebe_2016}, quantum amplitude estimation \cite{ramoa2024}, Mach-Zehnder interferometry \cite{pezze2014} and photonic phase estimation \cite{lumino2018}.

These scenarios consider a two-level quantum system whose binary measurement probabilities undergo oscillations at an unknown frequency \cite{Ferrie_2012, Wiebe_2016, Granade_2012, Wiebe_2014a}, induced by the time-independent Hamiltonian:
\begin{equation}
    H = \frac{\omega}{2}\sigma_i,
\end{equation}

\noindent with $i \in \{x, y, z\}$. In the Bloch sphere, this causes a precession of the Bloch vector around the axis corresponding to $\sigma_i$.  For simplicity, we consider an observable $\sigma_j$, $j \neq i$ and an initial state $\lvert \psi_0 \rangle$ corresponding to the $+1$ eigenstate of this observable. We denote the measurement outcome corresponding to $\lvert \psi_0 \rangle$ by $x=0$. The resulting outcome distribution is then given by: 
\begin{equation}
    \label{eq:likelihood}
    \mathbf{P}(x\mid \omega;t)=\left(\sin^2(\frac{\omega t}{2}) \right)^x\left(\cos^2(\frac{\omega t}{2}) \right)^{1-x},
\end{equation}

\noindent with $x \in \{0,1\}$ a binary outcome, $\omega$ the parameter of interest (a frequency to be estimated), and $t$ the evolution time before measurement. The latter being a controllable parameter, it can be used as an experimental control; measuring the system at different times amounts to sampling from different instances of a parametrized probability distribution. Our goal is to choose a sequence of measurement times, optimizing the data collection of $(t,x)$ tuples to enable quantum-enhanced estimation of $\omega$. 

Bayesian inference is a widely adopted technique for this task \cite{qinfer, Granade_2012, Wiebe_2016, Santagati_2019, Joas2021, Gentile_2021, Flynn_2022, fioroni2025, Arshad2024, Lumino_2018, mcmichael2021}. It is capable of real time estimation, can incorporate pre-existing information, and naturally provides estimates of the uncertainty, as well as assessments of the merit of generative models. It also admits highly parallelizable numerical treatment \cite{doucet2013}.

Additionally, the Bayesian framework is capable of leveraging the degree of freedom given by experimental controls such as $t$ to optimize the data collection process. In quantum metrology, this enables quantum-enhanced, and even Heisenberg-limited, estimation \cite{Higgins_2007, ramoa2024}. 

Bayesian inference relies on the systematic application of Bayes' rule:

\begin{equation}
    \label{eq:bayes}
    \mathbf{P}(\omega \mid x; t) = \frac{\mathbf{L}(\omega \mid x; t)\mathbf{P}(\omega)}{\mathbf{P}(x ; t)},
\end{equation}

\noindent which shapes our beliefs based on experimental observations. These beliefs are represented by a probability distribution over $\omega$, that quantifies the merits of any value it can take. Previous knowledge, or lack thereof, is represented by the prior distribution $\mathbf{P}(\omega)$. Those initial values are then weighted according to how likely each parameter would have been to generate the observations, quantified by the likelihood $\mathbf{L}(\omega \mid x; t) = \mathbf{P}(x\mid \omega;t)$ given by equation \ref{eq:likelihood}. The  denominator $\mathbf{P}(x ; t)$ is the marginal likelihood or model evidence; for our purposes, it can be regarded as a normalizing constant. 

Experimental data - measurement outcomes for various evolution times, or tuples $(t,x)$ - can then be processed using equation \ref{eq:bayes}, producing a posterior distribution $\mathbf{P}(\omega \mid x; t)$ that better aligns with the empirical reality. 

Estimates of $\omega$ and the associated uncertainty can be extracted from the posterior by integrating over it to get averages. Furthermore, one may use these distributions to evaluate utility functions for any given control $t$ (or sequence thereof) in a look-ahead fashion. One may then optimize measurement times to, for instance, minimize the variance. This is called Bayesian experimental design \cite{rainforth2024}. While the cost of this optimization is exponential in the number of experiments, this can be curbed using a locally optimal strategy, which has been shown  (numerically) to produce good results \cite{ferrie2012, msc}. 

In this greedy approach, only one experimental control is optimized at once. The corresponding measurement is then performed, the Bayesian distribution updated, and only after is the following control optimized, based on the updated distribution. Pseudo-code for a greedy Bayesian algorithm is given in algorithm \ref{alg:inference}. This offers two key advantages: reduced optimization costs, and the ability to leverage all available information at each point in time.  

\input{sections/pseudocode}

However, the optimization cost is still high: calculating the expected utility for a single prospective experimental control $t$ in one iteration has a cost over twice as high as the baseline cost of the iteration. Evaluating one utility requires performing two hypothetical Bayesian updates (one for each binary outcome), integrating the utility function over the two resulting distributions, and integrating the likelihood twice over the original distribution (to weight the utilities based on the probability of each scenario). This cost is then multiplied by the number of evaluations necessary for the optimization. 

In light of this, heuristics have been proposed to cut down the optimization costs. They differ in the \texttt{CHOOSE\_EVOLUTION\_TIME} sub-routine of algorithm \ref{alg:inference}. One such proposal is the $\/\sigma$ heuristic (SH) \cite{Ferrie_2012}, which is based on a theoretical argument that locates the optimum for the greedy strategy under a normality assumption. This suggests that the best pick is the measurement time $t = 1/\sigma$, where $\sigma$ is the standard deviation of the latest Bayesian distribution. This choice has the same worst case scenario as any other, but a better best case scenario. 

One related heuristic is the particle guess heuristic (PGH) \cite{Wiebe_2014a}, which leverages the numerical representation of the probability distributions to avoid the analytical calculation of $\sigma$. Instead of $\sigma$, a proxy with a similar expected value but additional statistical variability is employed: two samples are drawn from the distribution, and their distance computed to replace $\sigma$.

These heuristics have been widely adopted in the literature, including in noisy scenarios \cite{qinfer, Granade_2012, Wiebe_2016, Santagati_2019, Joas2021, Gentile_2021, Flynn_2022, fioroni2025, Arshad2024, Lumino_2018}. While they may perform well for lightweight strategies, their behavior can be unreliable, as the underlying arguments are based on assumptions of normality that do not always hold.  Furthermore, they are not necessarily optimal, nor do they offer a way to achieve optimality or otherwise improve the performance. Similarly, they are ill-suited to estimation under finite coherence, given that they rely on ever-increasing (indeed, exponentially increasing \cite{ferrie2012, msc}) measurement times. 

We note while phase estimation is a similar problem in its iterative formulation \cite{kitaev1995, Kitaev_2002, Wiebe_2016}, as the parametrized measurement circuit produces a likelihood similar to equation \ref{eq:likelihood}, circuit-based techniques for Fourier transform based phase estimation, such as \cite{van_Dam_2007}, no longer directly apply; we are dealing with a two-level system rather than a quantum gate that we can incorporate in a circuit. Other related works have considered frequency and phase estimation in the presence of noise, but focus on state preparation rather than adaptive feedback for the measurement times \cite{Fr_wis_2014, Macieszczak_2014, Demkowicz2011, Górecka_2018, Friis_2017}.

In this work we propose and numerically validate two novel frequency estimation algorithms, inspired by a recently proposed quantum amplitude estimation algorithm \cite{ramoa2024}. They are based on a window expansion strategy (WES) that drives an adaptive problem-tailored definition of the search range. These methods allow extra classical resources to be traded in for increased quantum advantage. 

One of the methods seeks variance minimization, while the other considers only a measure of statistical efficiency, making it more general and robust - for instance, it works under multi-modality, unlike standard deviation based methods (as are both the heuristics). We call these methods WES and annealed WES, or aWES (after the annealed sampling method it is inspired by, \cite{neal1998}), respectively. We additionally employ additional cost-cutting measures to lighten the classical processing load, prioritizing the application of resources to the algorithm's key components while sparing them in less impactful sub-routines.  

To ensure generality, we assess our algorithms' performance using randomized frequencies. As a benchmark, we use the comparable state of the art, namely the $\sigma$ and particle guess heuristics. We additionally compare these 4 approaches with a reference measurement strategy that uses random times (RTS), and locate them relative to the fundamental limits of metrology. We show that our algorithms are more reliable than any other, achieving the fastest and most stable learning rates.

In addition to ideal simulations, we test these methods under decoherence, studying how their performance deviates from ideal-case behavior. We show that in this noisy case, the improvements in performance brought by our algorithms is even more significant. 

The rest of this paper is organized as follows. Section \ref{sec:wes} describes the WES and aWES algorithms. Section \ref{sec:methods} details the methods used in the simulations and for benchmarking. Section \ref{sec:results} presents and discusses the results of the numerical simulations performed for all the considered frequency estimation strategies: WES, aWES, SH, PGH and RTS. Finally, section \ref{sec:discussion} summarizes the findings and points directions for future research.  

%% file: sections/pseudocode.tex
\begin{figure*}
    \centering
    \begin{minipage}{0.9\textwidth}
        \begin{algorithm}[H]
        \caption{Algorithm for greedy Bayesian frequency estimation. The algorithms analyzed in this work follow this structure, differing in the \texttt{CHOOSE\_EVOLUTION\_TIME} sub-routine. The specification of \Call{Bayesian\_update}{} is determined by the chosen numerical integration method; see \cite{ramoa2024} for an example.}
        \label{alg:inference}
            \begin{algorithmic}[1]
            \Require   \texttt{prior}($\omega$) \Comment{Distribution encoding prior knowledge on $\omega$.}
            \Ensure ${\hat \omega}$ \Comment{Estimate of the frequency.}
            \State \Getst {\texttt{d}($\omega$)} \texttt{prior}($\omega$)
            \While {not  \ttt{termination\_criterion}} \Comment{Maximum evolution time/uncertainty.}
            \State \Getst t \Call{choose\_evolution\_time}{\ttt{d}($\omega$)} 
            \State \Gets {x} \Callt{measure}{t} 
            \State \Getst {d($\omega$)} \Callt{Bayesian\_update}{d($\omega$), t, x} \Comment{Equation \ref{eq:bayes}.}
            \EndWhile
            \State \Gets {\hat \omega} \Callt{mean}{d($\omega$)}
            \State \Return ${\hat \omega}$ 
            \end{algorithmic}
        \end{algorithm}
\end{minipage}
\end{figure*}

%% file: sections/wes.tex
\section{Window expansion strategy}
\label{sec:wes}

This section presents the key insight behind the window expansion strategy (WES), based on \cite{ramoa2024}. Pseudo-code capturing the main idea is provided in algorithm \ref{alg:wes}. The involved constants were chosen heuristically, and shown to perform universally well (for arbitrary frequency values). However, they can be changed if one wishes to do so; we observed robust performance for a wide range of values. 

\input{sections/pseudocode_wes}

The likelihood in equation \ref{eq:likelihood} is oscillatory with frequency proportional to $t$. Low evolution times bring reliable but less informative data; longer times bring more peaked likelihoods but also complications, namely redundancy \cite{msc}. Thinking of normal distributions as an example, short times bring broad distributions, which are reliable but have large uncertainty; while long times bring distributions with narrower but not unique peaks (mixture of Gaussians in the domain), and this multi-modality brings ambiguity. Frequency estimation thus benefits from increasing times, bringing higher frequency likelihoods as knowledge accumulates. If low frequency information has already narrowed down the range of potential values for $\omega$ in parameter space, we are more likely  to overcome the multi-modality of a high frequency datum, by identifying the correct mode. 

More specifically, for this type of problem, the optimal evolution times scale exponentially with the iteration number \cite{ferrie2012}. However, the exact rate of this increase is execution-dependent, due to the stochastic nature of the estimation process. As such, we opt for an adaptive strategy that tailors the search range for the experimental controls adaptively, depending on the progress of the inference.

To start, we make $10$ measurements for $t=1$. This ``warm up'' with a low evolution time is due to the fact that the accuracy of the utility evaluation strongly depends on our knowledge, which is non-existent at this point (uniform prior). Thus, we need not spend resources in optimization for such small returns. Note that the units of time and frequency are relative, as only their product shows up in equation \ref{eq:likelihood}. We consider $\omega \in ]0, \pi/2]$; in general, this domain is problem-dependent. 

After these initial measurements, the distribution is typically close to normally distributed. We can then start optimizing based on it. We start with a small search range up to $t_\text{max}=100$. This is the initial \textit{window}. For the first iterations, we evaluate the utilities for $50$ possible controls sampled uniformly at random in this range. These evaluations can be performed in parallel. Note that while we suggest particular constants that work robustly for randomized test instances, WES can be adapted to better suit the problem at hand. For instance, one may make the optimization more thorough by increasing the number of utility evaluations; this should lead to lower estimation errors by improving the learning rate, at the expense of higher classical cost.

As a utility function to maximize, we use the additive inverse of the expected variance\footnote{In theory, we would expect to minimize ${\sigma^2*\text{CET}^2}$, with CET the cumulative evolution time; however, interestingly, we find that this results in a higher classical processing cost for negligible returns, due to low evolution times being favored.}. Expected values can be calculated using integration over the Bayesian distributions, just as parameter or uncertainty estimates, to be discussed below. The control with highest utility defines the control for the next measurement. We consider single-shot measurements in the optimization, to keep costs down. However, we use $10$ shots for the measurement, intuitively expecting that the best single-shot control will be a good choice for 10 shots as well. In practice, we find this has good results, while reducing processing costs and allowing parallel measurements (if we can measure identical systems governed by the same Hamiltonian) or a reduction in online processing (i.e. more measurements can be performed sequentially, without classical computations in between).  

The choice of the control is also used to inform the window adaptations. If the chosen control was one of the $3$ highest (among the $50$ possibilities), we count a \textit{hit} and proceed to the next iteration. Once $3$ hits (not necessarily consecutive) occur, a window expansion takes place: the upper bound of the search range becomes the new lower bound, while the upper bound doubles. The hit counter is then reset to $0$, and the process starts again. 

The intuition is that the selection of high controls signals that the learning process is advanced enough to benefit from high evolution times. However, this is a trend and not a certainty, due to the randomness of the process. We do not want to be precipitated in expanding the window, as choosing larger controls than the inference is ready for brings poor performance. This is why we wait until $3$ hits have accumulated to act, smoothing over stochastic effects. Likewise, numerical artifacts may cause the highest control not to be chosen,  and thus we consider a range of $3$.

Importantly, even as the window expands and its width repeatedly doubles, the number of utility evaluations is fixed at $50$. This ensures that the cost of optimizing the control does not increase with the iteration number. Intuitively, we expect the granularity in the optimization to be less critical as the evolution times grow; the effect of choosing time $t=1$ vs. $t=5$ should differ more than that of $t=1001$ vs. $t=1005$. Still, the number of utility evaluations can be increased to enhance performance, at the cost of additional classical processing. The ability to navigate this trade-off is an advantage over SH and PGH.

Further details of our algorithms are tied with the numerical integration method, which realizes the representation of the Bayesian distributions and update rules. We use sequential Monte Carlo (SMC) with a normal Metropolis resampling kernel \cite{doucet2013, south2019, daviet2016, msc, granade2012}. SMC is essentially a dynamic grid that traverses a sequence of distributions, periodically refocusing on the relevant regions of parameter space via a variety introducing mechanism, called the resampling kernel, with good convergence properties. We use Markov Chain Monte Carlo resampling. For pseudo-code and details, including on the expected utility calculations, refer to \cite{ramoa2024}.

In our case, the distributions in the sequence are the consecutive Bayesian probability distributions resulting from the cumulative datasets.  The resampling kernel refreshes the grid point locations when necessary - this necessity being diagnosed by the effective sample size (ESS), a measure of statistical efficiency. Below-threshold values signify poor representation and trigger a resampling step.  The ESS can be straightforwardly estimated by the SMC representation \cite{ramoa2024}.  

In WES, we skip the resampling in the utility calculations. In other words, we are less careful with the statistical representation when integrating to optimize the experimental controls (estimating utilities) than when integrating to get our final results (estimating the parameter). This does not jeopardize correctness and has a limited impact on performance, as the unitary look-ahead does not result in drastic deviations from the original distribution. A slightly inaccurate representation leads in the worst case to slightly sub-optimal choices; in practice, we do not observe a significant impact on performance, despite the substantial reduction of the runtime.  This greatly reduces the cost overhead associated with optimization.

\subsection{Annealed variant}

Annealed or tempered importance sampling \cite{neal1998, south2019} is an SMC algorithm for sampling from a complex distribution. In this case the distributions in the sequence are increasing powers of the target distribution, with exponents growing from $0$ to $1$. This effectively raises the posterior up slowly from a flat distribution, giving the resampling time to act when necessary. The successive powers are chosen to keep the effective sample size (ESS) stable around a target value, which is key to ensure good results.

We adapt this idea to our case, leveraging the structure of the likelihood function. Larger controls take the role of higher coefficients, as they are similarly more informative but also more challenging to sample from. We thus choose at each iteration the control that keeps the effective sample size the closest to a target value of $50\%$, which is a standard choice \cite{Doucet_2009, granade2012}.

Note that the sample size does not concern the uncertainty, information gain, or any such metric of success; it merely provides information about the correlation between samples, which affects the quality of the representation. However, the statistical representation has a great impact in all components of the inference process. Furthermore, we expect that keeping it around a chosen value will choose good controls: neither prematurely high, because that would lower the ESS too much, nor unnecessarily low, as that would keep the ESS too high. 

Thus, the annealed version of our algorithm, aWES, consists of an identical framework to WES, but with an ESS based utility replacing the variance minimization. The window expansion considerations still apply, as this quantity is to be evaluated for hypothetical controls as any other utility function. 

\subsection{Cost analysis}
\label{sub:cost}

We now analyze the classical processing cost of our algorithm and others, and specify in what sense our approach may be termed a ``low-cost'' one. We consider the cost in terms of elementary operations (multiplication or addition). The final calculations are presented in table \ref{tab:cost}. Details on their derivation are given in Appendix~\ref{app:cost}.

\input{sections/table_cost}

Global optimization is clearly problematic, with a cost exponential in the number of experiments. Additionally, it does not offer a straightforward way to define the search range, or choose $M$. Local optimization removes the exponential cost problem, but not the other. WES (or aWES) solves both, and additionally lowers the cost. In the tests we performed, the average total number of contemplated controls (potential evolution times) per execution, $M$, exceeded $10^8$; which would clearly entail a prohibitive cost if using local optimization. Using a dynamic search range allows WES to keep the number of candidate controls per iteration very low as compared to a static approach. 

Finally, the SH has a cost roughly $18$ times smaller than WES. This difference is expected, as these heuristics are lightweight rules of thumb that barely require calculations. The particle guess heuristic is even more parsimonious, with a further $4$-fold reduction in cost. However, performance deteriorates with these cost reductions. The goal of WES is to offer a systematic way of improving results, as heuristics do not, without incurring the hefty costs of traditional Bayesian experimental design; hence the epithet of \textit{low-cost Bayesian experimental design}. The numerical results of section \ref{sec:results} show that WES maintains an optimal learning rate, scaling at the Heisenberg limit; this suggests that the strategic resource savings employed therein do not impair performance as compared with full blown optimization, unlike heuristics such as the SH and PGH. 

It should be noted that these costs are given as a function of the total number of experiments $N$. In practice, this is not all that matters, as it fails to consider the ultimate metric of success: the estimation error. An algorithm may have a higher cost per experiment, yet compensate by learning more from each experiment. For instance, WES halves the number of experiments as compared to SH, and reduces the error (section \ref{sec:results}); this brings its cost closer to that of the SH. In contrast, RTS relies on an over $100$ increase in the number of experiments $N$ as compared to any other approach, and this too while achieving an error over $10$ times larger; negating the economy that table \ref{tab:cost} may naïvely seem to suggest.  Similarly, concerns tied with the stability of the numerical representation may affect the practical consequences of these costs.

%% file: sections/pseudocode_wes.tex
\begin{figure*}
    \centering
    \begin{minipage}{0.9\textwidth}
        \begin{algorithm}[H]
\caption{Window expansion strategies (WES and aWES) for optimizing evolution times in greedy Bayesian frequency estimation. Expansions of the search range are triggered by long evolution times. The initialize function is called once before the loop of algorithm \ref{alg:inference}, defining initial constants. As with \Call{Bayesian\_update}{} of algorithm \ref{alg:inference}, the specification of \Call{expectation}{}, \Call{variance}{} and \Call{ess}{} is determined by the chosen numerical integration method; see \cite{ramoa2024} for an example.}\label{alg:wes}
\begin{algorithmic}[1]  
\Function{initialize()}{}
        \State \textit{global } \texttt{hits}, $t_\ttt{min}$, $t_\ttt{max}$ \Comment{Shared state across function calls.}
        \State \Getst {hits} {$0$}
        \State \Gets {t_\ttt{min}}{$0$}
        \State \Gets {t_\ttt{max}} {$100$}
\EndFunction
\vspace{.5em}
\Require \ttt{d}($\omega$), \texttt{annealed} \Comment{Latest Bayesian distribution, and whether to use aWES.} 
\Ensure $\ttt{t}_\ttt{opt}$ \Comment{Optimized evolution time.}
\Function{choose\_evolution\_time}{\ttt{d}($\omega$)}
    \State \textit{global } \texttt{hits}, $t_\ttt{min}$, $t_\ttt{max}$ 
    \If {$\texttt{hits} ==3$}  \Comment{Trigger window expansion.}
        \State \Gets {t_\ttt{min}}{$t_\ttt{max}$}
        \State \Gets {t_\ttt{max}}{$2*t_\ttt{max}$}
        \State \Getst {hits} {$0$} \Comment{Reset hit counter.}
    \EndIf 
    \State \Getst {tlist} $50$ samples from \texttt{UNIFORM}[$t_\ttt{min}, t_\ttt{max}$]  \If {\texttt{annealed} is true then}
        \State \Getst {expectations} [\Call{expectation}{\Call{variance}{}, \ttt{d}($\omega$), \ttt{t}} for \ttt{t} in \ttt{tlist}]	
    \Else  
        \State \Getst {expectations} [\Call{absolute\_value}{(\Call{expectation}{\Call{ess}{}, \ttt{d}($\omega$)), \ttt{t}}$-0.5$} for \ttt{t} in \ttt{tlist}]	
    \EndIf 
    \State \Getst {$\ttt{t}_\ttt{opt}$} \Call{argmin}{\ttt{expectations}}
    \If{$t_\ttt{opt} \geq \Call{third\_highest\_value}{\ttt{grid}}$ }
        \State hits = hits + 1
    \EndIf
    \State \Return $\ttt{t}_\ttt{opt}$
\EndFunction
\end{algorithmic}
\end{algorithm}
\end{minipage}
\end{figure*}

%% file: sections/table_cost.tex
\begin{table}[ht]
\centering
\begin{tabular}{@{}lc@{}}
\toprule
\textbf{Strategy} & \textbf{Classical cost} \\
\midrule
No optimization/random    & $N*K$  \\
\midrule
Global optimization    & $(7M2^N+N)*K$ \\
\midrule
Local optimization    & $(14M+1)*N*K$ \\
\midrule
Window expansion (WES)    & $71N*K$ \\
\midrule
$\sigma$ heuristic (SH)    & $4N*K$ \\
\midrule
Particle guess heuristic  (PGH)  & $N*K$ \\
\bottomrule
\end{tabular}
\caption{Classical processing cost in terms of elementary operations for Bayesian inference using various experimental design optimization strategies. Local optimization considers a unitary look-ahead. $N$ is the number of experiments, $K$ is the number of samples for the numerical representation, and $M$ is the number of different controls (times) considered. $M$ is of the order of $10^8$ for the tests presented here.} 
\label{tab:cost}
\end{table}

%% file: sections/methods.tex
\section{Methods}
\label{sec:methods}

To assess the performance of the various experimental design strategies, we use them to estimate frequencies and compare the results. The code and datasets are available at \cite{Ramoa_2025_WES}.

The crucial benchmark takes the form of double logarithmic plots showing the evolution of the root mean squared error (RMSE) with the cumulative evolution time (CET). The latter is the sum of all evolution times up to the present iteration, i.e. the sum of all controls multiplied by the corresponding shot numbers. The resulting lines can be compared with the standard quantum and the Heisenberg limits (SQL and HL), for which the RMSE is expected to evolve as $\mathcal O (N^{-1/2})$ and $\mathcal O (N^{-1})$ respectively, translating into straight lines with slopes $-0.5$ and $-1$ in log-log scale.

We consider two test cases: one in ideal conditions, and one with a finite coherence time $T=500$ that exponentially damps the oscillations. In the decoherence case, all algorithms are informed of the noise model and coherence time. In both cases the data are generated via binomial sampling. 

To avoid biased assessments, we sample the frequencies uniformly at random from $]0, \pi/2]$, and consider a normalized version of the root mean squared error (RMSE) among $100$ runs with different frequencies.  The prior is taken to be a flat distribution in this domain. The upper bound on $\omega$ can be rescaled. Relatedly, we leave units unspecified, as only the relation between frequency and time matters.

The adaptive data cannot be averaged directly for particular values of the CET, due to the large variability in the values of $t$; no two executions follow the same pattern, as the measurements and thus the results are stochastic. To process the results for legibility, we use a bin-log-average-exp strategy introduced in \cite{ramoa2024}; this produces consistent results with individual runs and other processing methods.

For the representation of the Bayesian distributions, we use sequential Monte Carlo with Markov Chain Monte Carlo \cite{doucet2013, south2019, daviet2016, msc, granade2012}. The statistical backbone and hyperparameter configuration is held fixed across all the tested strategies. 

For the random strategy, the times are picked at random between $]0, C]$, with $C$ an optimized constant, and sorted in increasing order (this is a common trend across all strategies, and benefits the numerical representation \cite{msc}). In the presence of a finite coherence time $T_c$, we consider $C=T_c$. Likewise, we consider a proportionality constant in the $\sigma$ and PGH heuristics, which we optimize before running the simulations.

%% file: sections/simulation_results.tex
\section{Simulation results}
\label{sec:results}

The individual estimation processes for each of the four key strategies are shown in Figure~\ref{fig:ideal_singles}. 

\input{sections/ideal_singles}

Although the heuristics perform remarkably well given their simplicity, our strategies - WES and aWES - display the most regular behavior, and follow the Heisenberg limit reliably. aWES slightly less so: it shows slightly irregular behavior during the early stages, although it evolves in parallel to the Heisenberg limit for most of the process. 

The SH behaves somewhat erratically\footnote{The SH additionally has problematic runs where the learning stagnates (not shown due to unreasonable runtime).}.  This is problematic, as we generally may want to perform inference up to any given value of error; a robust strategy should have a consistent performance. On the other hand, the PGH shows a smoother but ultimately worse learning rate. It could be that the stochastic nature of the protocol alleviates the instability; or alternatively, this may stem from a positive correlation between slower learning rates and smooth behavior, as conservative measurement choices tend to produce both.

Table~\ref{tab:quantresults} shows quantitative results for the data, obtained by curve fitting. We also include the random strategy RTS as a point of comparison.

\input{sections/table}

Our algorithms achieve an exponent $-1.00$ as the Heisenberg limit, with WES having a better multiplier than aWES. The third best performing strategy is the SH, with an exponent $-0.86$, while the PGH comes after with $-0.66$ (but a lower constant factor). The worst performing strategy is RTS, at exponent $-0.44$ - slightly lower than the SQL. 

We also note that in addition to producing more reliable results, WES and aWES used a lower number of experiments as compared to SH. This can be seen in table \ref{tab:nexp}. The PGH uses even less experiments, but the achieves an error over $10$ times larger for the same cumulative evolution time. The random times strategy, predictably, uses a very large number of experiments, becoming very memory and resource intensive. 

\input{sections/table_nexp}

Figure \ref{fig:noisy_singles} shows how the performance suffers given a short coherence time.  All of the algorithms share the same Bayesian framework, which allows them to incorporate noise information into the likelihood, and are thus (in principle) capable of learning even if noise is present. Nevertheless, not all measurement time choices are made equal. For instance, measurement times well beyond the coherence time are uninformative; given a perfect generative model, the estimates are expected to be proper, but have large associated average errors and uncertainties. 

\input{sections/noisy_singles}

 As expected, we found that the optimal multiplying factors in the SH and PGH were significantly lower in this scenario. This is due to the fact that long evolution times are uninformative, as the system decays to the maximally mixed state. Likely due to this conservative choice, the SH displays more stable behavior. According to the discussion of section \ref{sec:wes}, lower value of the constant brings less ambitious and less challenging data.

WES has a stable behavior, but a visibly reduced learning rate as compared to the ideal case of figure \ref{fig:WES_ideal}. It seems to initially slightly surpass the SQL (albeit far from the HL) but soon becomes parallel to it. This is expected given the limited coherence. In this case, aWES seems to perform slightly better, albeit once again more irregularly. Quantitatively, it seems that the statistical efficiency (ESS) utility metric of aWES fares better in this noisy scenario.


Finally, Figure~\ref{fig:all} directly compares all of the strategies for the noiseless (Figure~\ref{fig:noiseless_all}) and noisy (Figure~\ref{fig:noisy_all}) scenarios. This allows a better assessment of the individual learning rates and the differences between them. 

\input{sections/all}

In both scenarios, the best performing strategy is WES. In the noiseless case, it both offers the most reliable performance and achieves the lowest error for any evolution time (and conversely, the lowest evolution time for any error). The SH and aWES almost match it in the later stages, but display more instability throughout - especially the former. aWES performs very close (and parallel) to it for most of the span of the graph, although with a slightly worse factor. The PGH has worse error than the random strategy. 

In the presence of noise, the performance of the SH and PGH is especially impacted. Indeed, while they are capable of learning, neither does better than the randomized approach. Even though the absorption of noise into the likelihood assures correctness and the best achievable performance given the heuristic formulae, it does not address the root cause of the problem: the strategies are inadequate for estimation in these conditions. One may then, if looking for a low cost strategy, adopt random times instead, as the cost is lower - no adaptivity is required, nor calculations of the standard deviation.

%% file: sections/ideal_singles.tex
\begin{figure*}[!ht]
\captionsetup[subfigure]{width=\linewidth}%
\centering
\begin{subfigure}{.45\textwidth}
  \centering
  \includegraphics[width=\textwidth]{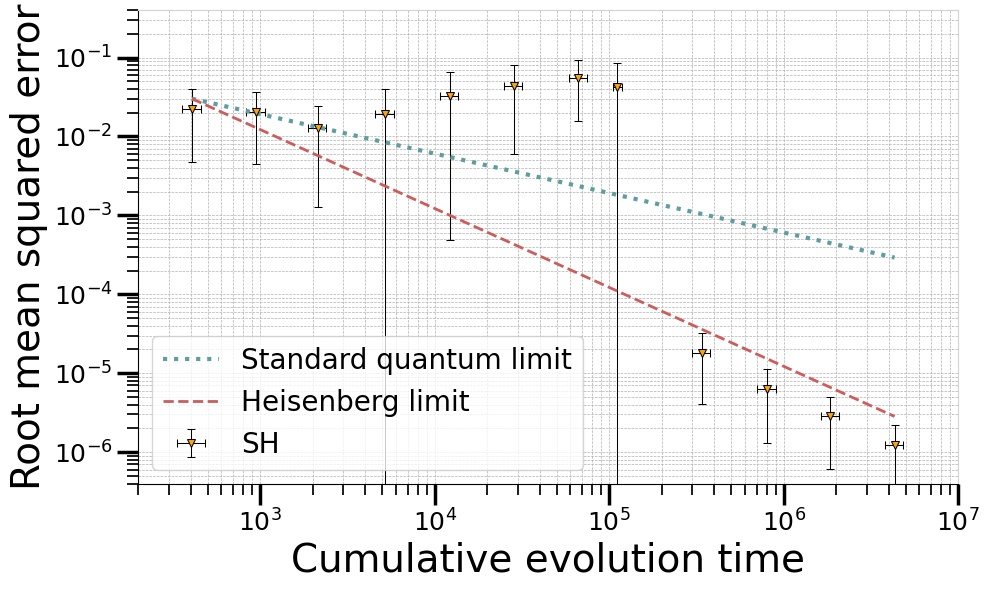}
  \caption{$1/\sigma$ heuristic (SH).}
  \label{fig:sigma_ideal}
\end{subfigure}
\hfill
\begin{subfigure}{.45\textwidth}
  \centering
  \includegraphics[width=\textwidth]{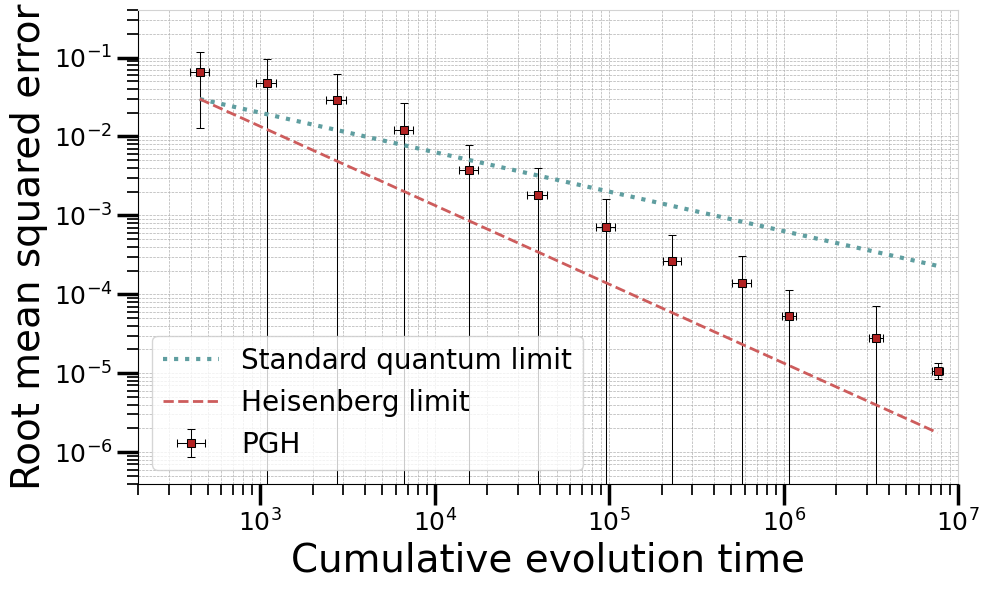}
  \caption{Particle guess heuristic (PGH).}
  \label{fig:PGH_ideal}
\end{subfigure}

\begin{subfigure}{.45\textwidth}
  \centering
  \includegraphics[width=\textwidth]{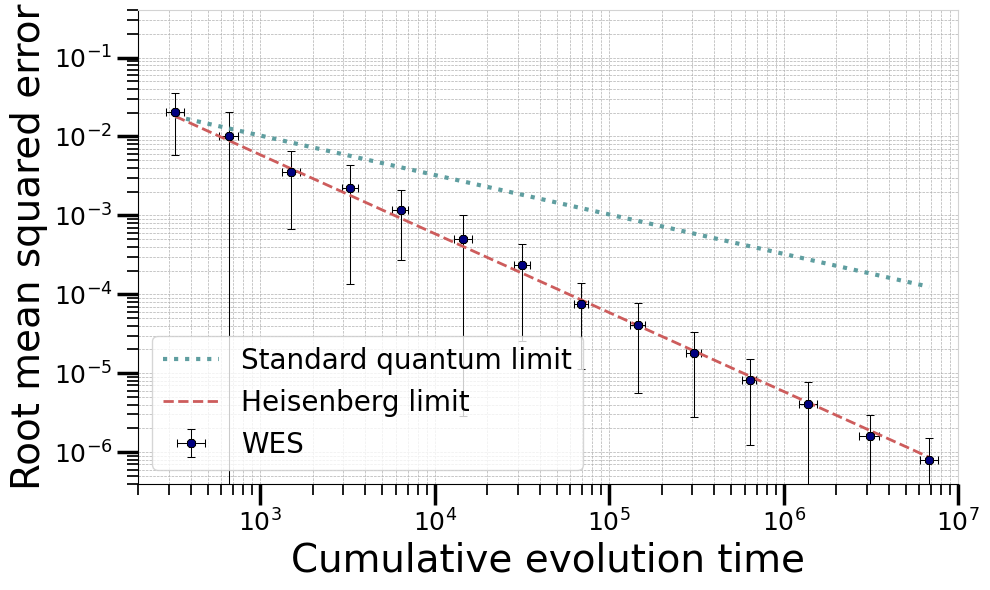}
  \caption{Window expansion strategy (WES).}
  \label{fig:WES_ideal}
\end{subfigure}
\hfill
\begin{subfigure}{.45\textwidth}
  \centering
  \includegraphics[width=\textwidth]{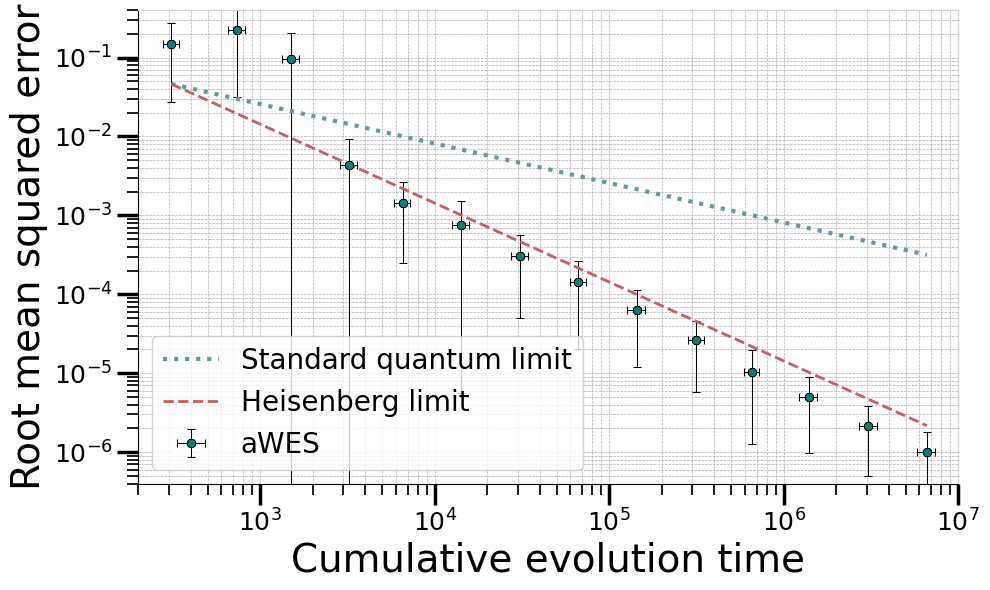}
  \caption{Annealed WES (aWES).}
  \label{fig:aWES_ideal}
\end{subfigure}

\caption{Evolution of the estimation error with the cumulative evolution time for each of the tested strategies under ideal conditions (no decoherence). Error bars represent one standard deviation. WES has the best and most reliable performance.}
\label{fig:ideal_singles}
\end{figure*}

%% file: sections/table.tex
\begin{table*}[h]
\centering
\begin{tabular}{@{}lSS@{}}
\toprule
\textbf{Strategy} & \textbf{Exponent/Slope} & \textbf{Factor/Offset} \\
\midrule
Window expansion strategy (WES)    & -1.00 & 1.78 \\
Annealed window expansion strategy (aWES)                & -1.00 & 2.66 \\
$\sigma$ heuristic (SH)             & -0.86 & 1.64 \\
Particle guess heuristic (PGH)     & -0.66 & 0.53 \\
Random times strategy (RTS)                            & -0.44 & -2.22 \\
\bottomrule
\end{tabular}
\caption{Fitted parameters for the exponent and factor (or slope and offset in log-log scales) of each experimental design strategy. The exponent/slope for the standard quantum and Heisenberg limits would be $-0.5$ and $-1$ respectively. WES and aWES saturate the Heisenberg limit.}
\label{tab:quantresults}
\end{table*}

%% file: sections/table_nexp.tex
\begin{table}[ht]
\centering
\begin{tabular}{@{}lc@{}}
\toprule
\textbf{Strategy} & \textbf{Experiments} \\
\midrule
Window expansion (WES)    & $3\times 10^2$ \\
Window expansion (aWES)    & $3\times 10^2$ \\
$\sigma$ heuristic (SH)   &  $6 \times 10^2$ \\
Particle guess (PGH)    &  $7\times 10^1$ \\
Random times (RTS)    &  $4\times 10^4$ \\
\bottomrule
\end{tabular}
\caption{Average number of experiments used by each strategy to achieve of results presented in figure \ref{fig:ideal_singles}. WES and aWES halve the number of experiments as compared to SH, while producing better results. PGH requires even less experiments, but the results are significantly worse. The random strategy requires a prohibitive number of experiments.} %
\label{tab:nexp}
\end{table}

%% file: sections/noisy_singles.tex
\begin{figure*}[!ht]
\captionsetup[subfigure]{width=\linewidth}%
\centering
\begin{subfigure}{.45\textwidth}
  \centering
  \includegraphics[width=\textwidth]{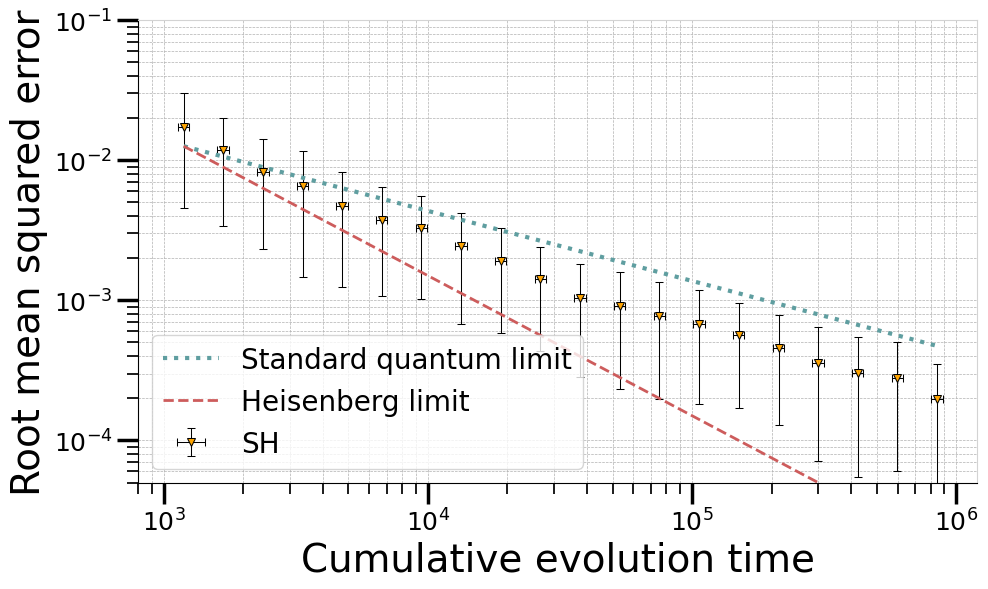}
  \caption{$1/\sigma$ heuristic (SH).}
  \label{fig:sigma_noisy}
\end{subfigure}
\hfill
\begin{subfigure}{.45\textwidth}
  \centering
  \includegraphics[width=\textwidth]{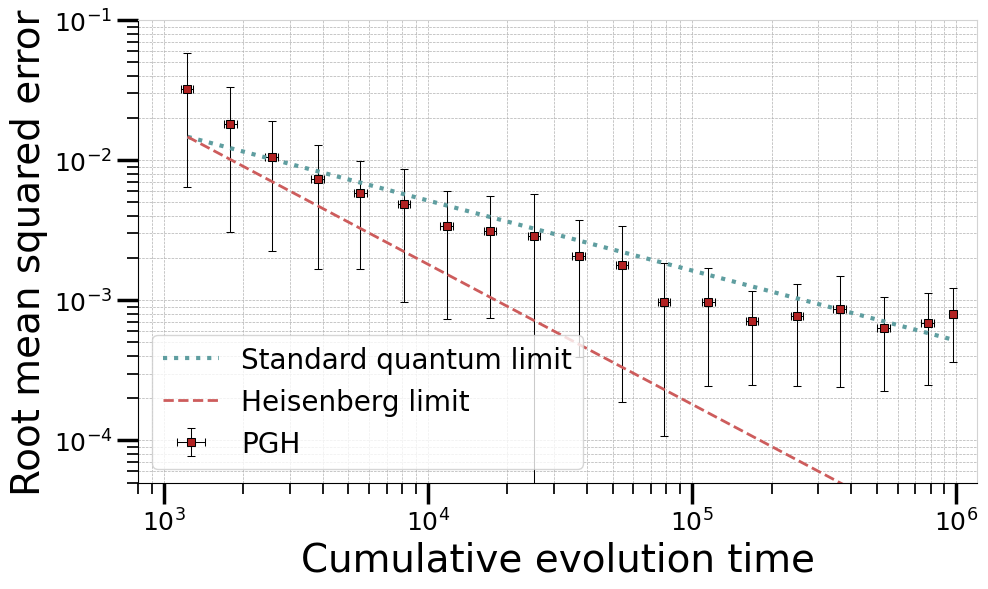}
  \caption{Particle guess heuristic (PGH).}
  \label{fig:PGH_noisy}
\end{subfigure}

\begin{subfigure}{.45\textwidth}
  \centering
  \includegraphics[width=\textwidth]{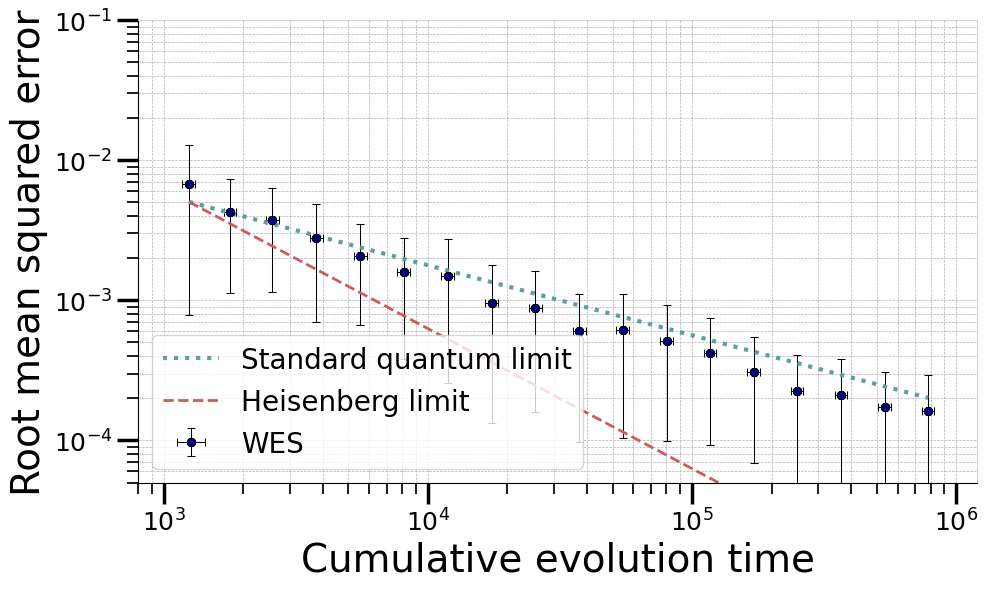}
  \caption{Window expansion strategy (WES).}
  \label{fig:WES_noisy}
\end{subfigure}
\hfill
\begin{subfigure}{.45\textwidth}
  \centering
  \includegraphics[width=\textwidth]{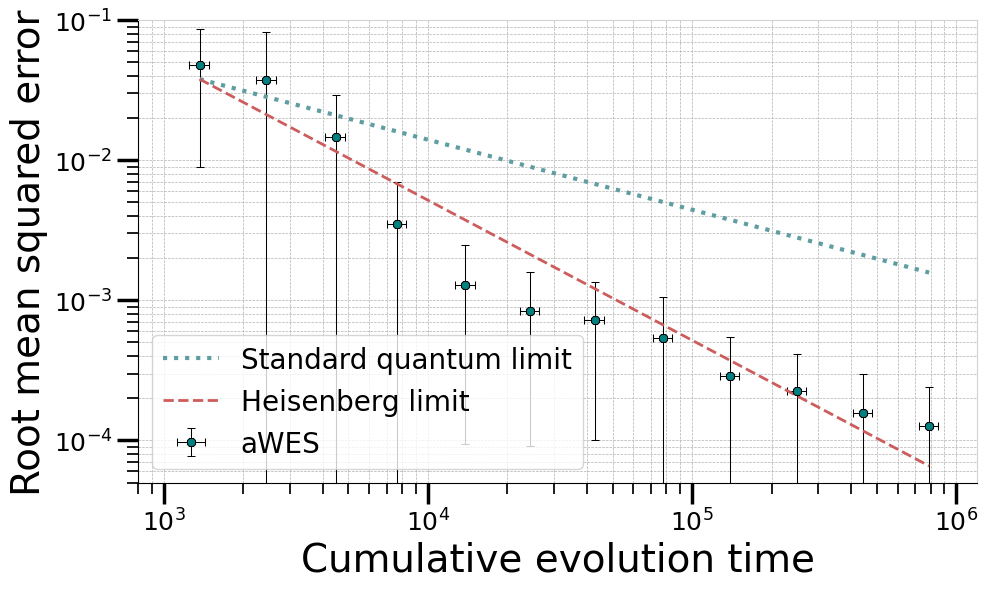}
  \caption{Annealed WES (aWES).}
  \label{fig:aWES_noisy}
\end{subfigure}

\caption{Evolution of the estimation error with the cumulative evolution time for each of the tested strategies under finite coherence ($T=500$). Error bars represent one standard deviation. WES and aWES achieve the lowest errors.}
\label{fig:noisy_singles}
\end{figure*}

%% file: sections/all.tex
\begin{figure*}[!ht]
\centering
\captionsetup[subfigure]{width=.9\textwidth}%
\begin{subfigure}{.5\textwidth}
  \centering
  \includegraphics[width=\textwidth]{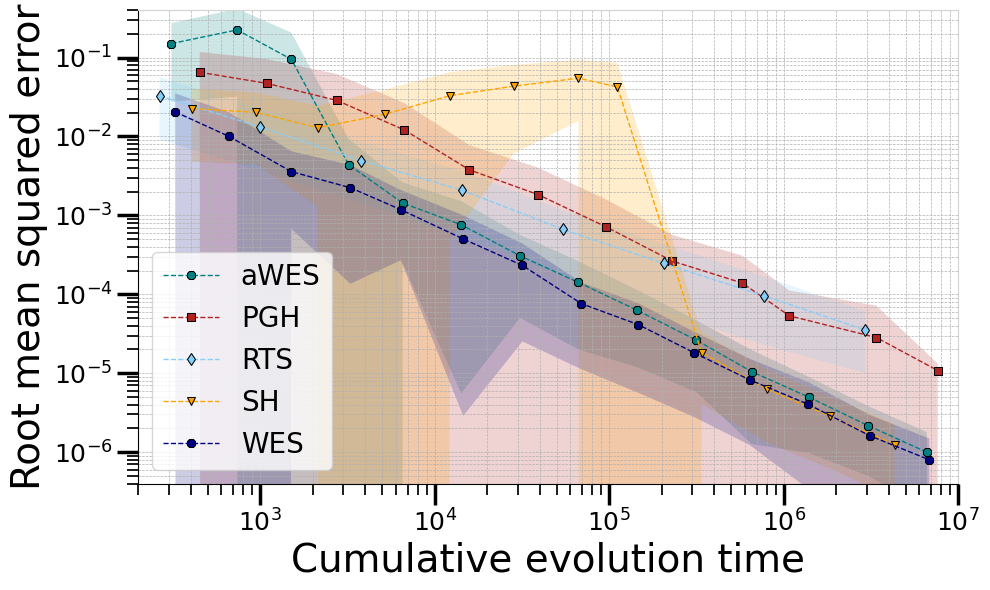}
  \caption{Ideal case.}
  \label{fig:noiseless_all}
\end{subfigure}%
\begin{subfigure}{.5\textwidth}
  \centering
  \includegraphics[width=\textwidth]{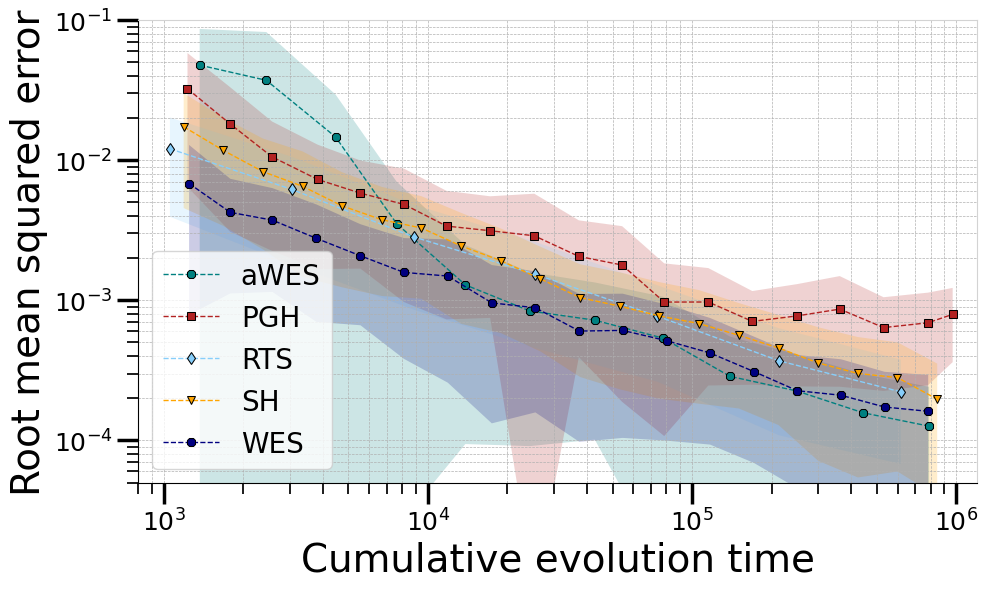}
  \caption{Finite coherence case. }
  \label{fig:noisy_all}
\end{subfigure}%
\caption{Evolution of the estimation error with the cumulative evolution time for all the tested strategies. Shaded areas represent the one standard deviation interval. WES and aWES are the best performing strategies.}
\label{fig:all}
\end{figure*}

%% file: sections/discussion.tex
\section{Discussion}
\label{sec:discussion}

We presented a comprehensive numerical benchmark for Bayesian frequency estimation techniques in the presence and absence of decoherence, testing commonly used heuristics as well as our own algorithms. The latter are based on a window expansion strategy (WES) that defines the search range for Bayesian experimental design on the fly depending on measurement results. In addition to a standard variance-minimization approach (WES), we develop an annealed variant (aWES) that optimizes a measure of statistical representativity.   

We identified shortcomings in standard heuristics, namely unstable behavior and unreliability in the presence of noise. Although these approaches are still relevant where a low optimization overhead is critical, their use may be counterproductive if the inference becomes inefficient due to poor experimental selection. In some cases, replacing the heuristics with random choices of measurement times produces as good or better results while reducing classical processing costs even further.

We found that WES outperforms all other algorithms in both noisy and noiseless scenarios, achieving Heisenberg-limited estimation. aWES performs almost as well as WES in the ideal case, and even better in the noisy case. This is remarkable for an algorithm that disregards all measures related to the uncertainty or information gain, focusing on statistical quality instead.  We expect this to have interesting ramifications for problems with more complicated (e.g. multi-parameter and/or multi-modal) likelihoods, as the numerical representation becomes more challenging. These tests are interesting directions for future research. 

Another such direction is to consider more sophisticated optimization methods. In this work, we have relied on brute force optimization for the Bayesian experimental design, for several reasons: it is easier to analyze, both analytically and numerically; and it removes confounding dependence on the specifics of the optimization. Nonetheless, similar insights to those introduced in WES could in principle be applied to general optimizers, to dynamically define the search range while adjusting the resources expended according to the phase of the inference process. It would be interesting to compare the cost-to-performance ratio of this WES-inspired optimization with optimization over a static search range. 

Further work could also extend this comparison to machine learning methods that have been gainfully employed for Bayesian experimental design in quantum problems \cite{Lumino_2018, Fiderer_2021}, and see how they compare to the strategies considered here.  

%% file: sections/akcnowledgements.tex
\section*{Acknowledgments}
This work is funded by national funds through FCT – Fundação para a Ciência e a Tecnologia, I.P., under the support UID/50014/2023 (https://doi.org/10.54499/UID/50014/2023), and supported by JST Moonshot R\&D Program Grant Number JPMJMS226C. A. R. acknowledges financial support by National Funds through the Portuguese funding agency, FCT - Fundação para a Ciência e a Tecnologia, within project LA/P/0063/2020 and under PhD grant 2022.12332.BD; and from the European Union’s Horizon Europe research and innovation program under EPIQUE Project GA No. 101135288.  

%% file: sections/app_cost.tex
\section{Estimates of the classical cost}
\label{app:cost}

We now detail the reasoning behind the estimates given for the classical processing costs in section \ref{sub:cost}.

Let $C_1$ denote the cost of one Bayesian update; $C_2$ the cost of computing an expected value; $K$ the number of samples used to represent the probability distribution; $M$ the number of possible controls; and $N$ the number of experiments, assumed to have binary outcomes. 

We have that $C_1 \approx K$, disconsidering normalization (which is not necessary until expectations are to be evaluated): $K$ products are necessary to update the probabilities as per equation \ref{eq:bayes}. On the other hand, $C_2 \approx 3K$: integrating requires weighting the target function by the distribution samples, summing the results, and normalizing. 

In the unoptimized case, where the controls are pre-determined (for instance, at random), the cost is simply that of $N$ Bayesian updates: $N*K$. If employing Bayesian experimental design, one must consider the cost of utility evaluations. This cost has two components: computing the conditional utility for a specific scenario, and computing the expected probability that such a scenario shall happen. The former has cost $C_1+C_2$, as one must perform a Bayesian update to get the conditional distribution and then integrate the utility function over it; the later is an integral of the likelihood, with cost $C_2$. The combined cost is then $C_1+2C_2 \approx 7K$, to be multiplied by the number of possible scenarios. 

If seeking the globally optimal strategy, the controls are optimized at once for the sequence of $N$ experiments. There being $M$ controls, and $2^N$ possible scenarios, the cost of utility evaluations is $7M*2^NK$. Adding the cost of the inference process itself, we get $(7M2^N+N)*K$ as in table \ref{tab:cost}.

The greedy strategy, which optimizes a single experiment at once based on the latest distribution, removes the exponential dependence on $N$. In this case, the resources spent in utility evaluations are $2M(C_1+2C_2)$ in each of the $N$ iterations, as there are only $2$ possible scenarios. The total cost becomes $(14M+1)*NK$.

For WES, we contemplate $50$ controls per iteration, and again $2$ possible scenarios for each, as WES also takes a greedy approach. The total cost would be $710N*K$ if performing single-shot measurements; however, as described in \ref{sec:wes}, we measure each optimized time 10 times, bringing a ten-fold reduction in cost per measurement. Therefore, the cost is $71N*K$.

As for the SH, which forgoes utility evaluations, the cost dedicated to choosing experimental controls is merely that of computing an integral (the standard deviation).   Thus, the total cost is $4N*K$. Finally, for the PGH, the cost of experimental choices is negligible; instead of integrating, two samples from the distribution are produced, and the inverse of their absolute difference calculated. The cost is approximately the same as that of the non-optimized case. 